# Reimagining Doctoral Training in Statistics: Is There a Role for a Professional Doctorate?

Camden L. Lopez[a]*

[a] *Department of Biostatistics, School of Public Health, University of Michigan, Ann Arbor, MI 48109 (Email: cllopez@umich.edu). The author thanks Philip S. Boonstra for helpful comments.*

* Corresponding author

# Reimagining Doctoral Training in Statistics: Is There a Role for a Professional Doctorate?

Modern demands of the statistics profession call for reimagining statistics training. The discipline needs to attract and develop students who are effective as real-world problem solvers, interdisciplinary collaborators, communicators, leaders, and teachers. Demand for statistics professionals with broad technical and non-technical skills has grown in a variety of settings, but especially in business and industry. Academic curricula, though, remain primarily oriented around a narrow, technical conception of statistics. Advanced graduate-level training essentially is limited to research doctorate (PhD) programs which tend to prioritize theoretical and methodological research over development of effective applied statisticians. Other professions, such as those of physicians and surgeons, have training oriented around a professional doctorate, as opposed to a research doctorate. The statistics profession should consider not only changes to PhD curricula, but also the potential for a professional doctorate, drawing ideas from the curricula of other professional degrees such as the MD.

## 1. Introduction

An effective system for training new statisticians is critical to the health of the statistics profession. Now that statistical methods pervade so much of science, industry, and government, often under the name of data science, the stakes for statistics training perhaps are higher than ever. Other disciplines such as computer science bring alternative approaches to data science, provoking calls for statisticians to rally to the challenge and reimagine the curricula for training new generations (He et al. 2019). There is a sense of opportunity for the statistics discipline, along with some discontent about its current state (Brown and Kass 2009) and concern about the consequences if the profession fails to meet modern challenges (Meng 2009). The degree to which the profession continues to mature and adapt will depend partly on how much it invests in maintaining an effective training pipeline, particularly in university graduate programs.

To date, graduate training in statistics and biostatistics essentially has been limited to master's and research doctorate (PhD) degrees. The PhD curriculum carries a heavy burden. Not only does it need to produce effective researchers who contribute to the discipline by developing new statistical theory and methodology or exploring new applications of existing methods; PhD graduates also are expected to be authorities in everyday applications of statistics which, though often having no need for a new statistical method, can have considerable complexity, whether for statistical reasons or otherwise. In many cases, a PhD is considered necessary for positions of responsibility or leadership in applied statistics. In clinical trials, for example, MS-level statisticians often have roles implementing analyses and reports, while PhD-level statisticians often are expected to be the source of decisions or guidance about what exactly to implement. PhD curricula and learning objectives reflect the statistician's dual role of researcher-practitioner, but research tends to have priority — the PhD is a research doctorate, after all. When it comes to updating statistics training at the doctoral level, there have been proposals and efforts to better prepare PhD students for collaborative, applied statistical roles. One wonders, though, whether applied statistics ever will get the widespread attention it deserves in a research doctorate setting; is that asking too much? Might there be a role in the future of statistics training for an entirely new kind of doctorate — one focused on extensive, practical competence and professional development in real-world applied statistics?

The purpose of this article is to raise the idea of a professional doctorate degree in statistics. What I envision as a professional doctorate is a degree certifying that the graduate (1) has a foundation of extensive exposure to statistical methods in one or more domains, including substantial real-world (not only classroom-based) experiences and practical skills in those settings; (2) has demonstrated ambition and capacity for holding a position of responsibility and leadership; and (3) will maintain certain professional standards. Whereas a PhD student is trained to carry out academic research building, refining, and exploring new uses for statistical tools, a professional doctorate student would be trained to skillfully and responsibly guide the use of those tools in real-world settings.

This idea is motivated, first, by a feeling that the current degree options in statistics and biostatistics give students too little opportunity to develop understanding, motivation, and skills for the real statistical needs of scientists, businesses, and government agencies. Second, the current options provide an unclear path for students who have the aptitude for higher levels of challenge, responsibility, and competence in their educations and careers, but have little interest in theoretical or methodological research. The idea for a professional doctorate also comes from a look at the training systems for other professions. In particular, I consider the training of medical doctors. Much of this article comes from my perspective as a biostatistician who frequently works with MD investigators, but I have assumed that most of the arguments apply to the broader discipline of statistics.

Opening a new channel of practice-focused doctoral training in statistics would require discussion, investigation, and planning beyond my knowledge and the topics covered in this article. My purpose is only to bring to the conversation about statistics training an idea that, to my knowledge, has not yet appeared in the literature. The following sections begin with some background, reviewing practical motivations (section 2) and idealistic motivations (section 3) for reforming statistics training. I comment on the current graduate degree options (section 4) before pointing to ideas from doctoral training in medicine (section 5) and giving a sketch of the potential curriculum for a professional doctorate in statistics (section 6).

## 2. Practical Demands for the Profession

Among practical concerns expressed in the literature about the training of statisticians, there have been several recurring themes. Some of these already have motivated changes in graduate program curricula. More changes are likely to come. It may be worth considering the potential for a professional doctorate to help address these issues.

One concern is that statistics students need more, and earlier, experience using statistical tools to solve real problems. For example, Zelen (2003, pp. 3427–8) expressed the opinion that "[t]he biostatistical scientist must know how science is conducted and be a participant," and that "it is important in the pre-doctoral period for the student to obtain meaningful experiences in the applications of biostatistics to real problems." Brown and Kass (2009, p. 108) also favored early exposure to cross-disciplinary research and real-world problem solving — partly to develop skills and interest in such work, and partly to make statistics programs more attractive to students driven by real-world applications. He et al. (2019, p. 5), summarizing the "Statistics at a Crossroads" workshop, identified "[t]he central role of practice" as a top priority for the discipline:

> Today it is imperative for us to put practice at the center of our discipline with relevant computation and theory as supports. Research and education in statistics and data science must aim at solving real world problems and must evolve with science and domain problems in general, with measurable impacts and contributions to the fields outside Statistics.

Preparing statistics students particularly for interdisciplinary work is a related concern (Begg and Vaughan 2011). Statistics departments might better prepare their students for working with non-statistical collaborators by providing explicit training and encouraging students to develop knowledge outside of statistics (DeMets et al. 1994, DeMets et al. 2006, Brown and Kass 2009). Graduate programs often require elective courses in areas of application, but the statistics faculty do not necessarily show serious interest in how students fulfill that requirement or integrate it with their other training. Lehoczky (1995) described the interdisciplinary training in the statistics MS and PhD programs at Carnegie Mellon University and emphasized the importance of the faculty being fully committed to the value of interdisciplinary work.

Some commentators would like to see statisticians take a more expansive view of their discipline and their roles, becoming more proactive and willing to assume leadership (Brown and Kass 2009, Hoerl and Snee 2010). Such a change in attitude likely would help improve the profession's poor visibility, which has been lamented repeatedly (Iman 1995, Kettenring 1997, Gibson 2019).

Communication and other non-technical skills frequently are a concern (Lachenbruch 2009, Gibson 2019). Pomann et al. (2021) neatly summarized both the technical and non-technical skills needed by collaborative biostatisticians in clinical/translational research which likely are relevant to statisticians in many work environments.

There has been a perception, going back decades, that the supply of statisticians has not kept up with demand (Utts 2015, p. 102). DeMets et al. (1994) noted the growing relevance of, and demand for, biostatisticians that was not being accompanied by similar growth in the number of biostatistics graduates. Particularly, there did not appear to be great interest in biostatistics graduate training among domestic US students. A large portion of the demand for students and graduates was being filled by international students. More recently, the same concerns were reiterated, and the proliferation of genomic and other biological data needing statistical analysis may have exacerbated the shortage of skilled graduates (DeMets et al. 2006). In some cases, jobs that could be filled by graduates from statistics and biostatistics programs are filled instead by graduates from computer science, engineering, bioinformatics, or other fields with quantitative training.

There is now greater demand for statistics graduates in business and industry, particularly in data scientist roles. The Mathematical and Statistical Sciences Annual Survey of new doctorate recipients ("Doctoral Recipients" 2023) provides some data on the employment of statistics and biostatistics PhD graduates (Figure 1). The number of doctoral graduates being employed in business and industry more than doubled from 2010 to 2018 and was near 50% of graduates employed in the US for 2015–2018. The number employed in academic departments or research institutes slightly declined from 2010 to 2018, and most recently, employment in doctorate-granting departments of mathematics, statistics or biostatistics represented about 20% of graduates. Employment in other academic departments or research institutes represented another 20% or so.

Finally, there is always a need for more effective teachers of statistics. Undergraduate statistics education has received particular attention (Horton and Hardin 2015). There is widespread recognition in other disciplines that statistical ideas and skills are important, which places greater pressure on statisticians as educators; Meng (2009, p. 203) put it this way:

Figure 1. US employment of new doctorate recipients in statistics and biostatistics, 1996–2018. The data were obtained from annual reports of the Mathematical and Statistical Sciences Annual Survey published by the American Mathematical Society.

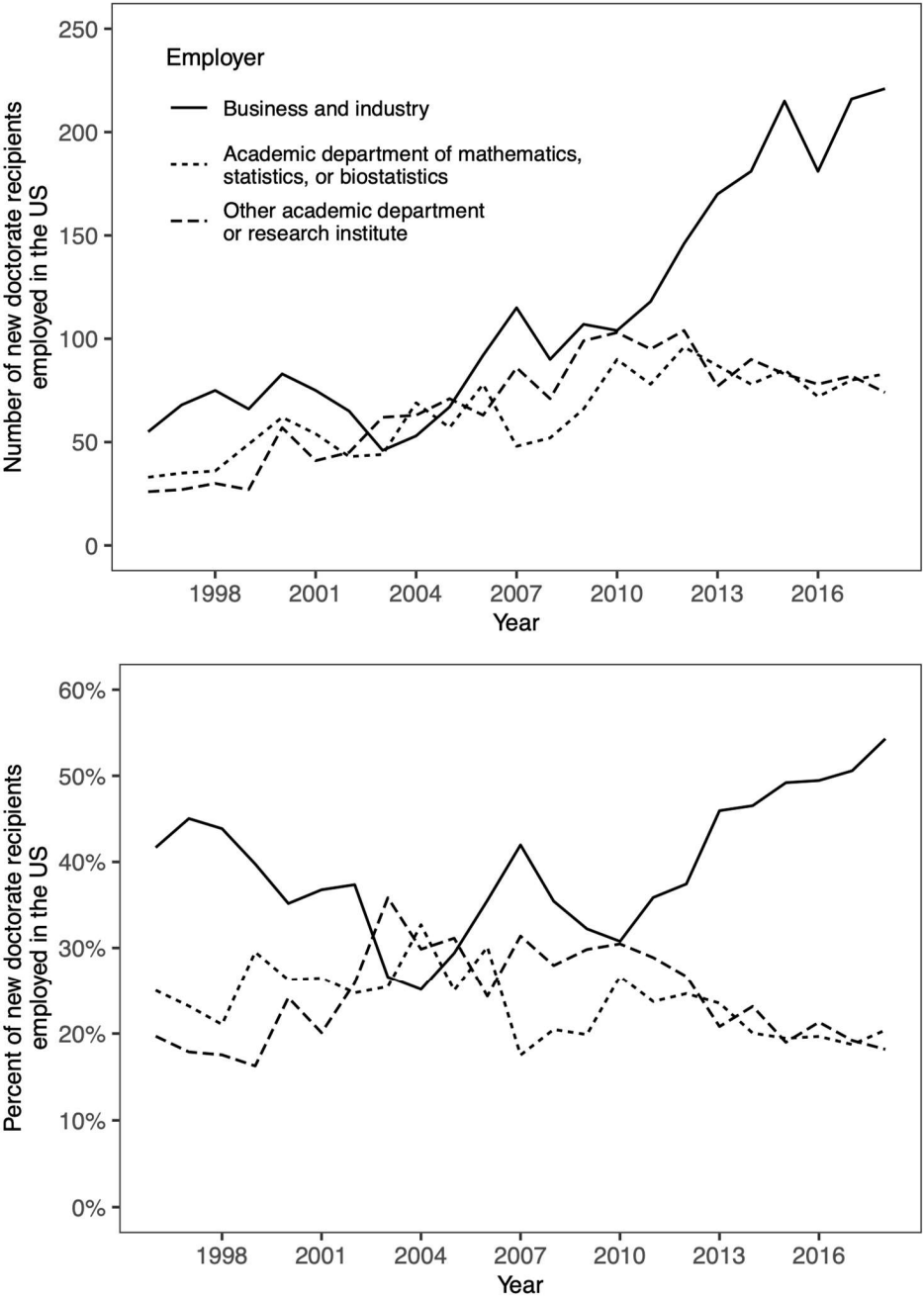

> *This should be our profession's deepest fear: we could screw up big time* because it is no longer just about helping others clean up their backyards, but rather about preparing whole generations of future scientists and policy makers.

Among the qualities of the ideal statistics educator are extensive practical experience and excellent communication. Greater emphasis on practical experience, communication, and leadership in statistics training could have the dual benefit of producing more effective applied statisticians and producing more effective teachers.

**3. Ideals for the Discipline of Statistics**

Although practical concerns are more likely to motivate academic departments to make curriculum changes, ideals might equally provide impetus for imagining how statistics training (particularly at the doctoral level) could be different. What should statistics be about? What should statisticians aspire to? The following two ideals are foundational for my view of graduate curricula today and the potential for a professional doctorate in the future. Although these ideals are unlikely to be controversial, I state them here because they were rarely, if ever, expressed in my own training.

First, statisticians ideally commit themselves to the actual problems of science (or medicine, industry, or government) — to developing theory and methods that start from and maintain contact with practical realities. Box (1976, pp. 797–798) emphasized the importance of feedback between theory and practice and the "perils of the open loop" — being "stuck either in the practice mode or in the theory mode." In academic statistics, the tendency is towards an excess of "theory mode." On the training of statisticians, Box wrote,

> A proper balance of theory and practice is needed and, most important, statisticians must learn how to be good scientists; a talent which has to be acquired by experience and example.

Additionally, when it comes to applied statistics work, I believe that the ideal statistician looks at scientific problems on their own terms, seeing them as worthy of attention primarily for the purpose of scientific or social benefit. In contrast, statisticians sometimes pay attention to real scientific problems because those problems provide ideas and data for theoretical or methodological research. Although this attitude may be necessary to some extent for statisticians whose careers depend on research, I believe there should be wariness towards it.

A second ideal, related to the first, is that statisticians should not accept a limited role for themselves. A statistician should not merely provide statistical solutions to satisfy the specific requests of an investigator, without questioning those requests or providing greater conceptual input. Bross (1974, p. 126) put it this way: "Am I a scientist or a shoe clerk?" The shoe clerk's responsibility is only to find an item in the store that pleases the customer. The scientist, in contrast, has broad responsibility towards public interests and professional principles.

This second ideal is related to the first because having credibility as more of a scientist than a shoe clerk requires knowledge about, and direct involvement in, the science itself — not the quantitative methods alone. As Bross (1974, p. 127) put it (pardon the non-inclusive pronouns), writing about statisticians in clinical trials and other public health research,

> If a statistician doesn't want to be treated as a shoe clerk, he has to show that he isn't one. If he wants to be treated as a professional he has got to act like a scientist and accept both the responsibilities and the hazards of a scientist in the public health area.

**4. Current Options for Graduate Training**

With the practical and idealistic concerns of the previous two sections in mind, do the current options for graduate training in statistics adequately support the needs and aspirations of the profession?

*4.1 Features of MS and PhD Curricula*

Graduate students and prospective graduate school applicants in statistics or biostatistics have essentially two options for the degree they pursue: MS (or other master's degrees; I will use "MS" as shorthand) or PhD. The MS curriculum typically includes (1) introductory probability and statistical theory, (2) basic methods and technical skills of applied statistics, (3) a structured, supervised experience in statistical consulting or data analysis for a real research problem, and (4) elective courses. The typical PhD curriculum adds (5) an additional year or two of coursework that includes measure theoretic probability and advanced statistical theory, plus advanced elective courses, and (7) a dissertation requiring around 2–3 years of research.

One feature of statistics MS and PhD curricula is that many of the student's early experiences and challenges are mathematical. Even in courses oriented around applied statistics, the more difficult and intellectually stimulating elements of the course material often have to do with the mathematical machinery of the statistical models and related computations. This machinery can be important for an applied statistician to understand. However, its prominence in the curriculum, especially early in the training of a statistician, should be considered in relation to the profession's goals for attracting students with a drive for real-world problem solving and an aptitude for communication and leadership. Early in training is when one begins forming an idea of what statistics is really about — what skills are valued, what opportunities there are for building on one's natural strengths, and the degree to which one has interest and aptitude for more advanced training. Much of the introductory coursework, to say nothing of the advanced coursework and dissertation research, will feel inviting to those with the most mathematical abilities while creating distress and discouragement for others. Meanwhile, deficiency in another area such as communication, leadership, or scientific curiosity is less likely to be an obstacle to progression through the graduate curriculum.

Another feature is that experience in the application of statistics in a more realistic context often is limited. For many students, their first such experience in consulting or analysis is one of the most impactful and memorable parts of their graduate program, so it is surprising that departments do not often offer additional, more advanced training. Although many PhD students get experience in applied statistics through work as a research assistant (RA), usually such work is part of a funding mechanism, not an explicit part of the curriculum. As a result, the student's RA experience is highly dependent on particular funding arrangements and the degree to which the faculty supervisor has the interest and ability to create experiences that help the student mature into a professional. Development and demonstration of one's consulting or collaboration skills could be made more explicitly a goal of doctoral training and a topic for more deliberate instruction.

The current PhD curriculum largely makes sense for the goal of producing a graduate who, at the completion of the curriculum, can develop a new statistical method or a novel application and successfully publish it in an academic journal. Statistical theory and methodology have developed to such an extent that several years of focused preparation really are necessary to have an adequate understanding of one corner of the literature, to say nothing of contributing something new.

Today, however, a large portion of PhD graduates will be employed in roles involving little, if any, academic statistical research. There is such extensive statistics literature that only a small portion of research and business projects need a truly novel algorithm or model. Understanding existing methods with enough breadth and depth to provide confident, competent guidance about their use — a critical function if statisticians are to be valuable and relevant — is no small challenge.

For those PhD students who will serve primarily as applied statisticians after graduation, the attention they must give to focused methodology research during their training years leaves them limited opportunity for developing a broader, more practical mental framework and skill set. Pocock (1995, pp. 212–213), commenting about the training of medical statisticians in the UK, wrote,

> Compared with other scientists, I think the medical statistician has a particularly marked shift of emphasis from theory to practice after his taught courses, which it is important to commit oneself to if one intends to become a useful collaborative scientist. At that stage, a theoretical PhD devoted to a particular methodological problem may actually delay true career development, especially if the PhD experience is largely divorced from the practical motivations of the methodology (an experience all too common in U.K. PhDs in statistics).

### *4.2 The Choice Between MS and PhD*

On the one hand, an MS entails a shorter time to complete the degree and minimal academic research, if any. Some practically minded students may choose this option out of a sense of fatigue from being a student or an eagerness for experience outside academia.

However, an MS graduate leaves with only limited exposure to the discipline. Employment opportunities often are limited to positions with lower levels of autonomy and responsibility. Years of experience may allow one to advance to higher levels, though some roles can be effectively, if not officially, closed to those without a PhD.

On the other hand, the PhD requires much more time, and the exact amount of time is uncertain. Persistence, discipline, and a tolerance for years of immersion in statistical literature and academic culture, often while having only limited contact with practical uses of statistics, are needed to complete the dissertation. The PhD qualifies one for employment that generally involves greater autonomy, greater responsibility, and the expectation that one will contribute not only technical skills but also ideas. Some students will seek a PhD because they aim for an academic career. Others will be drawn to the intellectual challenge or the social status accorded to those with doctorates. Others may proceed with a PhD only because it is what high-performing students are expected to do.

I hypothesize that many students would be attracted to a third option: doctoral training that focuses on the professional practice of applied statistics rather than methodology research. Among students who choose to obtain only an MS, I believe that a significant number would welcome an opportunity to develop practically relevant expertise and improve their prospects for more challenging, responsibility-laden careers. In the PhD curriculum and student experience, however, they see long years of lonely work that often appears unsatisfying and without much relevance outside a narrow corner of the academic literature. At the same time, among PhD students, there is a portion who have little interest in statistical methodology research *per se* and would be more interested in applying statistics to real-world problems. Yet they press on through their dissertation work — through a program designed, in large part, to train a student in research — because that stamp of the doctoral degree is needed for employment commensurate with their ambition and abilities. Granted, such students might still enjoy their PhD experience for the intellectual challenge, camaraderie, and personal growth; but the same enjoyment could happen in a program that leaves them with more relevant skills.

Some might argue that if one is interested primarily in being an applied statistician — if one has little interest in contributing to academic research within statistics — then graduate training is adequate as it is structured now: the MS gives you the formal education you need, and any additional knowledge or skills you need can be learned on the job. It may be true that work experience is adequate (in some ways it may be better than university training) for growing one's abilities for applied work. But this view ignores the way in which a student who chooses the MS path gets stamped as being less serious and less capable of contributing intellectually to a project. Some MS students feel like second-class citizens in their department, compared to PhD students; after graduation, this hierarchy remains in some workplaces, particularly academic institutions. The presence or absence of a "D" after one's name can significantly affect how one is perceived, what kinds of contributions to a project are expected or welcomed, and what opportunities one has to continue developing as a professional.

PhD students benefit from more attention from faculty and more of the kind of attention that one gives a potential colleague. The doctoral training experience is not only about developing technical expertise. It is a process of becoming socialized into a professional class — a process of developing a sense of confidence and professional identity while becoming connected to peers and senior figures in the profession. A professional doctorate in statistics would offer such benefits to a broader pool of students, particularly those who are highly driven but practically minded.

## 5. Ideas from the Training of Medical Doctors

Some professions are heavily oriented around developing new knowledge and methods — scientists and mathematicians, for example. The main path to enter such a profession is through a research doctorate, the PhD. Other professions are heavily oriented around using existing knowledge and methods in practice — lawyers and pharmacists, for example — and the training centers around a non-research doctorate, such as a JD or PharmD.

Professional statisticians are needed for research within the discipline (developing new statistical theory and methods), but also, ever increasingly, for practical application of the discipline's existing theory and methods. Another profession involving a balance of research and practice is medicine. In thinking about how to train statisticians to meet modern demands, what can statistics learn from medicine?

The MD degree forms the basis of training nearly all physicians and surgeons (the alternative being the DO degree). MD curricula generally include the following components: (1) coursework in basic science, organ systems, and diseases; (2) courses or seminars introducing students to the profession, describing the training expectations, and beginning to develop practical skills of interacting with patients; (3) clerkships that usually last a full year altogether and consist of rotation through various departments of a hospital, with the student being directly involved in patient care and exposed broadly to the practice of medicine in different settings; (4) more advanced coursework, often tailored to the student's career interests; (5) further clinical experience, such as a sub-internship, helping the student transition to residency; and (6) a scholarly research experience, usually lasting from several months to a year. Additional clinical training in the form of a residency, generally lasting at least three years, is required to become licensed to practice. Thus, for much of 5 or more years (2 in an MD program, 3 or more in residency), the trainee is immersed in patient care and first- or second-hand experiences of the responsibilities and skills of practicing doctors.

A few specific features of MD training, besides the immersive practical experiences, are worth noting. First, from early in medical school, developing a shared understanding of the medical profession and a professional identity is an explicit part of the program. A recent tradition is for the student's entry into the profession to be marked with a "white coat ceremony" and a pledge to uphold certain ideals (Weiner 2018). Some schools designate time in the curriculum for students to reflect on their experiences and develop their individual professional identity — for example, "Physician Identity (PI) Weeks" (2022) at University of California San Francisco. While many experienced statisticians likely have a strong sense of the profession's role and of how one forms and maintains an individual professional identity, such ideas could be communicated more consistently to students.

Second, MD training covers systematic, concrete skills for approaching patient interactions. In contrast, statistics trainees receive minimal systematic instruction about how to interact with collaborators or manage a project over time. MD training is further concerned with developing a student's clinical reasoning. Due to the complexity of both patients' conditions and the available treatments, clinical reasoning necessarily involves more than scientific knowledge. Analogously, applied statistical reasoning involves more than mathematical and theoretical statistics knowledge, but statistics trainees mostly are left on their own when it comes to developing a way to think through problems in applied statistics. Or worse, they are led to believe that the statistician only has a role after the investigator has framed the problem in such a clean, detailed manner that the statistician hardly needs any non-mathematical knowledge at all.

Third, in medical training there is some recognition that one's learning necessarily will continue after formal training ends, and that the state of medical knowledge may change dramatically over time. The medical profession takes measures to encourage its members to update and grow their knowledge throughout their careers (Duffy and Holmboe 2006). Charles Burwell, dean of Harvard Medical School, famously told medical students in 1944 ("Past Deans of the Faculty of Medicine" 2023),

> Half of what we are going to teach you is wrong, and half of it is right. Our problem is that we don't know which half is which.

In statistics graduate programs, what is being taught is not necessarily wrong, but it never can cover all useful topics, and some portion of it will become obsolete. Trainees, and perhaps even seasoned statisticians, could benefit from skills for assessing and updating their knowledge, and for finding well-justified guidance when faced with an unfamiliar statistical problem.

Statisticians might consider the high priority that medical training places on practical experience, even for students aiming to be medical researchers. Doctoral training in research alone is not considered sufficient. Those aiming for a research career can get additional preparation with an MD-PhD degree or by other means, such as fellowships, but foundational clinical proficiency is mandatory.

Efforts to update medical training (Skochelak and Stack 2017, Schwartzstein et al. 2020) might be a source of inspiration for updating statistics training. Among other changes occurring at many medical schools, clinical experience is beginning earlier so that classroom-based education is more integrated with clinical practice throughout the curriculum. Traditionally, the first two years of medical school were primarily coursework, and clinical experiences happened in the third and fourth years. Now many programs start clinical clerkships in the second year. A number of schools have implemented redesigned curricula recently (for example, University of California Los Angeles), and some are in the process of redesigning (for example, University of Pittsburgh).

Indeed, He et al. (2019, p. 25) noted the idea of "a 'medical school model' of core training followed by rotations" for statistics PhD curricula. The recently created Biomedical Data Science PhD program at University of Wisconsin–Madison includes three semester-long research rotations. Other statistics programs might also have elements that are inspired by, or analogous to, clerkships and professional development in MD training already.

To be clear, these observations about medical training are not meant to suggest that MD training is necessarily effective, or superior to statistics PhD training, in producing well-prepared professionals. The purpose is only to generate ideas for statistics training by looking at a profession that has a longer history and is known for producing large numbers of both practitioners and researchers.

**6. A Professional Doctorate in Statistics**

What might a professional doctorate in statistics look like? Practical experience should start early, and as much as possible, it should be integrated with more traditional classroom learning. Early practical experience (perhaps even in the first graduate year) could better motivate students in their theoretical-mathematical coursework. A type of rotation through immersive, prolonged, practical experiences in different settings of applied statistics could be considered. Some students already work as a statistical analyst for a year or two after completing their MS to obtain experience before enrolling in a PhD program; a year of employment as a student analyst-apprentice, arranged within a university or by cooperative arrangements with other employers, could be a formal part of a professional doctorate curriculum. Becoming a well-informed, effective member of a research or business team should be an explicit part of the training. The goal should be to produce a statistician who is ready, or nearly ready, to serve as a lead statistician and make relevant recommendations that are technically, practically, and ethically sound.

The training could embrace a wider domain of knowledge than statistics departments customarily embrace — not merely understanding probability, estimation, hypothesis testing, and prediction, but also understanding (or at least being aware of) concepts in the scientific or medical domain; issues around publication, funding sources, budgeting, and data management; management of junior statistical workers; and regulatory and ethical issues. The student could be given a sense of the societal role of statisticians, typical career paths, and how one develops a professional identity. Perhaps most importantly, the principal values, challenges, and aspirations of the profession could be communicated to students early and consistently.

Development of a professional doctorate might be accompanied by an expansion of what is considered statistical methodology, and by greater development of a literature around the "theory of applied statistics." Greater attention could be given to the close examination of case studies, as Birnbaum (1971) suggested. The idea of statistical engineering (Hoerl and Snee 2010) is an example of this type of broader thinking about statistical theory and methodology. The reasoning involved in tackling real problems in statistical science could be further studied and systematized.

If a professional doctorate — I will provisionally call it an SD degree, for *Statisticae Doctor* (my best attempt at Latin) — were implemented, the MS and PhD degrees likely would continue to be relevant. How would the SD fit in with those degree options? One possibility might be that students choose between the SD and PhD relatively early, and that the two curricula diverge (though with considerable overlap in coursework and other elements) after the student completes the MS curriculum. Another possibility is that the SD eventually serves as the basis for all doctoral statistics training, in a manner similar to how MD/DO degrees are the basis for all doctoral training in medicine. All doctoral students in statistics could have a similar SD curriculum which is augmented for those students wishing to train more in theory and methodology research, either by obtaining an SD-PhD, or perhaps by completing one or more postdoctoral fellowships.

Table 1 sketches what elements the MS, SD, and PhD curricula might have in common and in distinction from one another. This is not intended to be a proposal of a specific curriculum plan, but only a starting point for further discussion. Regardless of the exact structure of the programs, it would be desirable for students and faculty more interested in practical applications to remain in close contact with those more interested in theory and new methodology, and vice-versa.

A professional doctorate could have a well-defined schedule of completion, likely three or four years. The PhD — because of the demands of original research, variation in what is judged a sufficient amount of work to complete the dissertation, and variation in how much faculty advisors push students to complete — necessarily has a longer time to degree and significant uncertainty about exactly how long it will take, which can be a source of distress for students. A fully structured, scheduled doctorate program could be an attractive alternative.

Table 1. A sketch of potential curricula for three degree options in statistics and biostatistics, including a professional doctorate.

| Component of Curriculum | Degree Option | | |
| --- | --- | --- | --- |
| | Master's (MS) | Professional Doctorate (SD) | Research Doctorate (PhD) |
| Introduction to the profession of statistics<br>*Degree options, expectations, professional roles and ideals* | ✓ | ✓ | ✓ |
| Introductory statistical theory, methods, and computation | ✓ | ✓ | ✓ |
| Introductory experience in consulting and data analysis | ✓ | ✓ | ✓ |
| Elective courses in statistical or computational methods | ✓ | ✓ | ✓ |
| Courses in a scientific discipline or other application area | ✓ | ✓ | ✓ |
| Introductory professional skills<br>*Presentations and reports, skills for lifelong learning* | ✓ | ✓ | ✓ |
| Immersive experiences in multiple domains of statistical science<br>*Similar to internships or apprenticeships, analogous to MD clerkship rotations* | | ✓ | |
| Mini-courses or seminars about various contextual topics<br>*Regulatory issues, data sharing/standards, equitable use of data and models* | | ✓ | |
| Introductory experience in statistical methodology research<br>*Lasting several months, potentially forming part of a dissertation* | | ✓ | ✓ |
| Advanced professional skills<br>*Dependent on career goals, potentially focused on academia or industry* | | ✓ | ✓ |
| Advanced statistical theory | | | ✓ |
| Additional experience in statistical methodology research<br>*Research sufficient to complete a dissertation* | | | ✓ |

## 7. Conclusion

Does a professional doctorate have a role in the future of statistics training? I see several potential benefits to speculate about. More students, particularly domestic students in the US, might become attracted to statistics generally, and doctoral training particularly, if there is a degree option that is welcoming to students motivated more by real-world problem solving than academic research. Professional doctorate curricula could give more attention to non-technical skills and extra-disciplinary knowledge, producing more effective collaborators, communicators, leaders, and teachers. Employers could benefit from a greater supply of practical problem-solvers in data science. Academic departments could focus their PhD programs on developing competence in theoretical and methodological research, rather than trying to satisfy demands for both research and practical skills in their PhD curricula.

I imagine that developing professional doctorate programs would encourage academic departments to recruit more faculty with experience and interests outside of academic statistics, perhaps including some who have trained or led research in other disciplines. Developing and executing a professional doctorate curriculum might also involve close cooperation with nearby employers of statisticians and data scientists. The attention of academic departments might be pushed towards real-world needs. Thereby, the ideals of statisticians being fully in touch with practical realities, and being true scientists rather than shoe clerks, might find greater expression in academic departments.

As it stands, these benefits are only speculative. The hypothesis that a professional doctorate would appeal to many students is based on my limited experience. Input from students, faculty, and employers is needed to better understand the potential. Data would be helpful. Others have remarked on how the statistics profession, ironically, does a poor job of collecting data on its supply of students, demands of employers, or the effectiveness of training (DeMets et al. 1994, He et al. 2019). Here is another opportunity for gathering and analyzing data on the profession itself. The problem of better understanding the student population's needs and outcomes resembles a common task in clinical biostatistics: identifying patient subtypes, describing prognoses, and targeting treatments to improve outcomes.

      Modern demands for the statistics profession call for reimagining statistics training. If statistics professionals are to have leading roles in the data science era, a professional doctorate might play a part in training those leaders.

# References

Begg, M. D. and Vaughan, R. D. (2011), "Are Biostatistics Students Prepared to Succeed in the Era of Interdisciplinary Science? (And How Will We Know?)" *The American Statistician*, 65(2), 71–79.

Birnbaum, A. (1971), "A Perspective for Strengthening Scholarship in Statistics," *The American Statistician*, 25(3), 14–17.

Box, G. E. P. (1976), "Science and Statistics," *Journal of the American Statistical Association*, 71(356), 791–799.

Bross, I. D. J. (1974), "The Role of the Statistician: Scientist or Shoe Clerk," *The American Statistician*, 28(4), 126–127.

Brown, E. N. and Kass, R. E. (2009), "What Is Statistics?" *The American Statistician*, 63(2), 105–111.

DeMets, D. L., Anbar, D., Fairweather, W., Louis, T. A., and O'Neill, R. T. (1994), "Training the Next Generation of Biostatisticians," *The American Statistician*, 48(4), 280–284.

DeMets, D. L., Stormo, G., Boehnke, M., Louis, T. A., Taylor, J., and Dixon, D. (2006), "Training the next generation of biostatisticians: A call to action in the U.S." *Statistics in Medicine*, 25, 3415–3429.

"Doctoral Recipients" (2023), American Mathematical Society, accessed January 2023. Available at https://www.ams.org/profession/data/annual-survey/phds-awarded.

Duffy, F. D. and Holmboe, E. S. (2006), "Self-assessment in Lifelong Learning and Improving Performance in Practice," *Journal of the American Medical Association*, 296(9), 1137–1139.

Gibson, E. W. (2019), "Leadership in Statistics: Increasing Our Value and Visibility," *The American Statistician*, 73(2), 109–116.

He, X., Madigan, D., Yu, B., and Wellner, J. (2019), "Statistics at a Crossroads: Who is for the Challenge?" National Science Foundation workshop report. Available at https://www.nsf.gov/mps/dms/documents/Statistics at a Crossroads Workshop Report 2019.pdf.

Hoerl, R. W. and Snee, R. D. (2010), "Moving the Statistics Profession Forward to the Next Level," *The American Statistician*, 64(1), 10–14.

Horton, N. J. and Hardin, J. S. (2015), "Teaching the Next Generation of Statistics Students to 'Think With Data': Special Issue on Statistics and the Undergraduate Curriculum," *The American Statistician*, 69(4), 259–265.

Iman, R. L. (1995), "New Paradigms for the Statistics Profession," *Journal of the American Statistical Association*, 90(429), 1–6.

Kettenring, J. R. (1997), "Shaping Statistics for Success in the 21st Century," *Journal of the American Statistical Association*, 92(440), 1229–1234.

Lachenbruch, P. A. (2009), "Communicating Statistics and Developing Professionals: The 2008 ASA Presidential Address," *Journal of the American Statistical Association*, 104(485), 1–4.

Lehoczky, J. (1995), "Modernizing Statistics Ph.D. Programs," *The American Statistician*, 49(1), 12–17.

Meng, X.-L. (2009), "Desired and Feared—What Do We Do Now and Over the Next 50 Years?" *The American Statistician*, 63(3), 202–210.

"Past Deans of the Faculty of Medicine" (2023), Harvard Medical School, accessed February 2023. Available at https://hms.harvard.edu/about-hms/office-dean/past-deans-faculty-medicine.

"Physician Identity (PI) Weeks" (2022), University of California San Francisco, accessed February 2023. Available at https://meded.ucsf.edu/md-program/current-students/curriculum/foundations-1/physician-identity-pi-weeks.

Pocock, S. J. (1995), "Life as an academic medical statistician and how to survive it," *Statistics in Medicine*, 14, 209–222.

Pomann, G.-M., Boulware, L. E., Cayetano, S. M., Desai, M., Enders, F. T., Gallis, J. A., Gelfond, J., Grambow, S. C., Hanlon, A. L., Hendrix, A., Kulkarni, P., Lapidus, J., Lee, H.-J., Mahnken, J. D., McKeel, J. P., Moen, R., Oster, R. A., Peskoe, S., Samsa, G., Stewart, T. G., Truong, T., Wruck, L., and Thomas, S. M. (2021), "Methods for training collaborative biostatisticians," *Journal of Clinical and Translational Science*, 5(1), 1–13.

Schwartzstein, R. M., Dienstag, J. L., King, R. W., Chang, B. S., Flanagan, J. G., Besche, H. C., Hoenig, M. P., Miloslavsky, E. M., Atkins, K. M., Puig, A., Cockrill, B. A., Wittels, K. A., Dalrymple, J. L., Gooding, H., Hirsch, D. A., Alexander, E. K.,

Fazio, S. B., and Hundert, E. M., for the Pathways Writing Group (2020), "The Harvard Medical School Pathways Curriculum: Reimagining Developmentally Appropriate Medical Education for Contemporary Learners," *Academic Medicine*, 95(11), 1687–1695.

Skochelak, S. E. and Stack, S. J. (2017), "Creating the Medical Schools of the Future," *Academic Medicine*, 92(1), 16–19.

Utts, J. (2015), "The Many Facets of Statistics Education: 175 Years of Common Themes," *The American Statistician*, 69(2), 100–107.

Weiner, S. (2018), "The solemn truth about medical oaths," AAMC, accessed February 2023. Available at https://www.aamc.org/news-insights/solemn-truth-about-medical-oaths.

Zelen, M. (2003), "The training of biostatistical scientists," *Statistics in Medicine*, 22, 3427–3430.